\documentclass[twocolumn,amsmath,amssymb,floatfix]{revtex4}

\usepackage{graphicx}
\usepackage{dcolumn}
\usepackage{bm}

\graphicspath{{Figures/}}
\setkeys{Gin}{width=\linewidth}

\begin{document}

\title{High-frequency gate manipulation of a bilayer graphene quantum dot}

\author{S. Dr\"oscher} \email{susanned@phys.ethz.ch}
\author{J. G\"uttinger}
\altaffiliation{Present address: CIN2 (ICN-CSIC), Catalan Institue of Nanotechnology, Campus de la UAB, 08193 Bellaterra (Barcelona), Spain}
\author{T. Mathis}
\author{B. Batlogg}
\author{T. Ihn}
\author{K. Ensslin}
\affiliation{Solid State Physics Laboratory, ETH Zurich, 8093 Zurich, Switzerland}

\begin{abstract}

We report transport data obtained for a double-gated bilayer graphene quantum dot. In Coulomb blockade measurements, the gate dielectric Cytop$^{\mathrm{TM}}$ is found to provide remarkable electronic stability even at cryogenic temperatures. Moreover, we demonstrate gate manipulation with square shaped voltage pulses at frequencies up to 100 MHz and show that the signal amplitude is not affected by the presence of the capacitively coupled back gate.
\end{abstract}

\maketitle


It has been proposed to use the spin of a confined electron as a two-level system representing a quantum bit \cite{Loss:1998}. A basic requirement for the controlled manipulation of individual spin states in such a qubit are long decoherence and relaxation times. Graphene is expected to meet this request due to the low interaction of the electron spin with the carbon host lattice \cite{Trauzettel:2007}. Recently, quantum confinement has been shown in graphene quantum dots \cite{Schnez:2009,Moser:2009,Liu:2010} and spin states could be identified \cite{Guttinger:2010}. However, spin relaxation times in graphene nanostructures remain to be determined experimentally at present.

Here, we demonstrate high-frequency (HF) gate manipulation of a double gated graphene quantum dot (QD). These experiments are a first step towards the read-out and control of the state dynamics of an electron residing in such a quantum system.

\begin{figure}
  \begin{center}
    \includegraphics{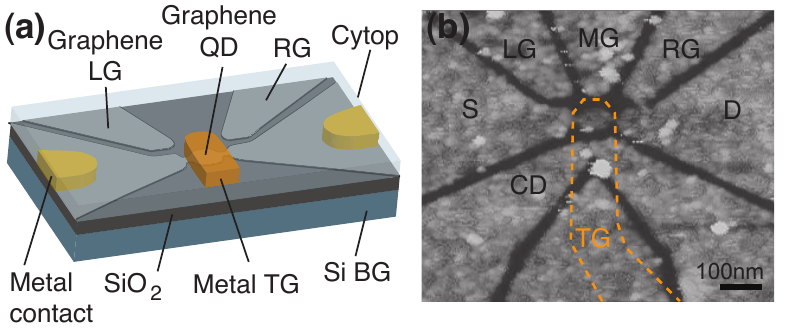}
    \caption{(a) Schematic of a top gated graphene quantum dot device. The graphene device (light grey) is contacted with gold electrodes (yellow) and covered a continuous layer of Cytop$^{\mathrm{TM}}$ (light blue). The metallic top gate finger (orange) covers the graphene quantum dot. (b) Atomic force microscope (AFM) image showing the sample before the top gate was processed. The island is located in the middle of the image and is connected to source (S) and drain (D). A number of gates (left gate (LG), middle gate (MG) and right gate (RG)) is located around the quantum dot and a charge detector (CD) is lying nearby. The area marked by the orange dashed line was in a second step covered by a top gate electrode (TG).}
    \label{fig1}
  \end{center}
\end{figure}

The QD device was fabricated from a bilayer graphene flake deposited onto Si/SiO$_{2}$ substrate following the process steps described in Ref. \onlinecite{Guttinger:2009}. The device layout is shown in the schematic (a) and the micrograph (b) in Fig.~\ref{fig1}. The graphene island has a size of 85$\times$50 nm$^2$ and the constrictions that constitute the tunneling barriers were measured to be 20 nm wide. The nearby charge detector is not used and connected to ground in the measurements discussed here.

The top gate (TG) dielectric used for this device is commercially available Cytop$^{\mathrm{TM}}$, a fluoropolymer used mainly for coatings and inorganic thin film field effect transistors \cite{Kalb:2007,Walser:2009}. It has a relative permittivity of $\epsilon \approx$ 2.1-2.2 and withstands high electric fields 
The chip was spin coated with Cytop CTL - 809 M from Asahi Glass Japan dissolved in CT - Solv. 180 in the ratio 1:10. Next, a local top gate was defined by electron beam (e-beam) lithography followed by metal evaporation (2 nm Ti/ 40 nm Au) and lift-off. Since Cytop$^{\mathrm{TM}}$ is highly water repellant, a 5 nm thick Cr-layer was evaporated prior to the application of the e-beam resist and removed in an etching step after patterning of the top gate. In Fig.~\ref{fig1} (b) the location of the TG electrode is depicted by the orange dashed contour.

Cytop$^{\mathrm{TM}}$ has previously been used as a dielectric at room temperature \cite{Kalb:2007,Walser:2009}. Experiments at 4 K were carried out in a dip stick setup to test the material stability at cryogenic temperatures. Standard lock-in techniques were applied to measure DC-transport through the quantum dot at constant voltage bias between source and drain reservoir. 

\begin{figure}
  \begin{center}
    \includegraphics{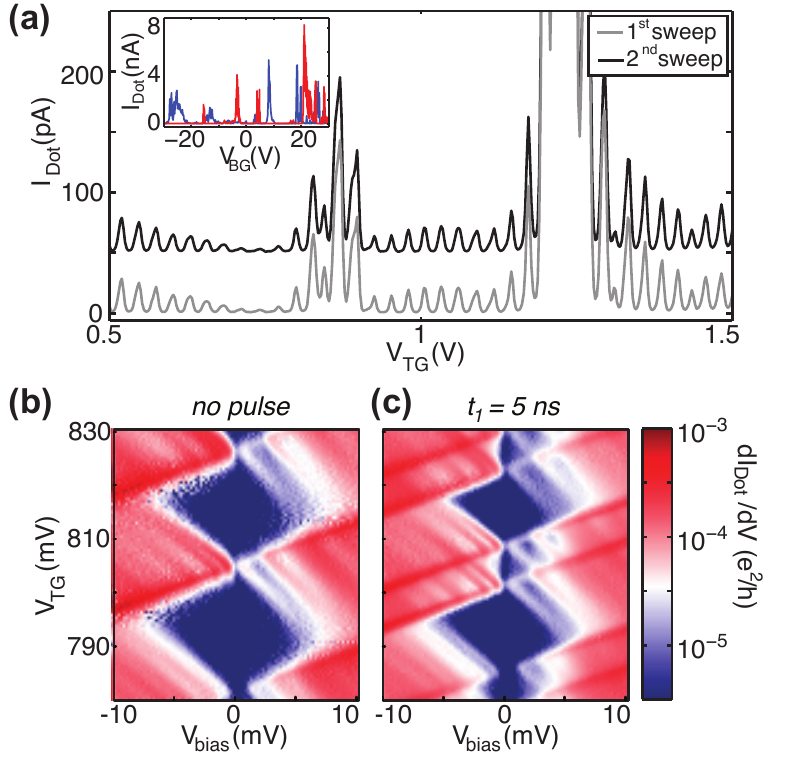}
    \caption{(a) Top gate dependence for $V_\mathrm{BG}$ = 0. The two sweeps were recorded at 4 K one after the other with $V_\mathrm{bias}$ = 5 mV. The black curve is shifted by +50 pA for clarity. Inset: Back gate characteristics before (blue trace) and after (red trace) the top gate was patterned. Both curves were recorded with a bias of $V_\mathrm{bias}$ = 5 mV at 120 mK and 4 K, respectively. (b) and (c): Finite bias spectroscopy at $T$ = 120 mK with (b) only DC-voltage at top gate and (c) AC-modulation with symmetricsquare pulses of $\Delta V$ = 100 mV amplitude and $f$ = 100 MHz frequency.}
    \label{fig2}
  \end{center}
\end{figure}

The recorded back gate (BG) dependence is presented in the inset of Fig.~\ref{fig2} (a). Within the measured gate range the current is mostly suppressed and only few spots exhibiting higher transmission are present. This behavior is qualitatively similar to the one observed in measurements taken on the same device before the TG was added, as shown by the blue trace in the same figure. In single layer graphene quantum dots, a region of suppressed conductance is commonly present in the back gate dependence \cite{Guttinger:2009} and is referred to as transport gap.

Zooming into the gapped region, but now sweeping the voltage applied to the top gate electrode while keeping $V_\mathrm{BG}$ = 0, illustrates the remarkable stability of the device. The two traces in the main figure of Fig.~\ref{fig2} (a), taken one after the other, fall perfectly on top of each other. They reveal a number of Coulomb blockade resonances that are mostly equally spaced as commonly observed for single quantum dots \cite{Kouwenhoven:1997}. 

From the size of the corresponding Coulomb blockade diamonds (see Fig.~\ref{fig2} (b)) a charging energy of $E_\mathrm{c} \approx$ 8~meV was extracted. Comparing the extracted island capacitance $C$, which is inversely proportional to $E_\mathrm{c}$, to the data obtained for single layer QDs, shows that the capacitance of the bilayer QD is significantly larger. This increased screening is a result of the additional metallic top gate electrode. Indeed, the maximum size of Coulomb blockade diamonds measured before adding the TG was $E_\mathrm{c} \approx$ 15 meV. This value is comparable to data obtained with single layer graphene quantum dots of similar size \cite{Guttinger:2009,Molitor:2009}.

Moreover, the evolution of Coulomb peaks was recorded as a function of an applied electric field applied normal to the bilayer sheet as well as for an external magnetic field (not shown here). Although a potential difference between the graphene layers was introduced by the  electric field, no band gap sizable on the scale of the charging energy was detected in the spectra. Further, the results obtained from the measurements in a magnetic field were found to be qualitatively different from the $B$-field characteristics exhibited by single layer QDs \cite{Guttinger:2009,Libisch:2010}. In the present device no dispersion or evolution towards Landau levels was observed, although predicted theoretically \cite{Libisch:2011}. Potential fluctuations in the bulk or introduced due to the edges of the structure are present and may cause this discrepancy. It remains to be shown whether this limitation is inherent for etched structures due to a dominating edge contribution to disorder or whether it can be overcome by decreasing the effect of the substrate on the bulk disorder.


We now proceed with the discussion of the pulsed gate experiments. All high-frequency measurements were carried out in a dilution refrigerator at a bath temperature of 120 mK. A bias tee was used to admix the pulsed signal to a DC offset. In this way both signals were applied simultaneously to the top gate. An arbitrary waveform generator (Tektronix AWG520) was used to generate the square shaped voltage pulses allowing for pulse lengths as short as 1 ns. Simultaneously, DC-transport measurements through the quantum dot were performed with symmetric voltage bias.

\begin{figure}
  \begin{center}
    \includegraphics{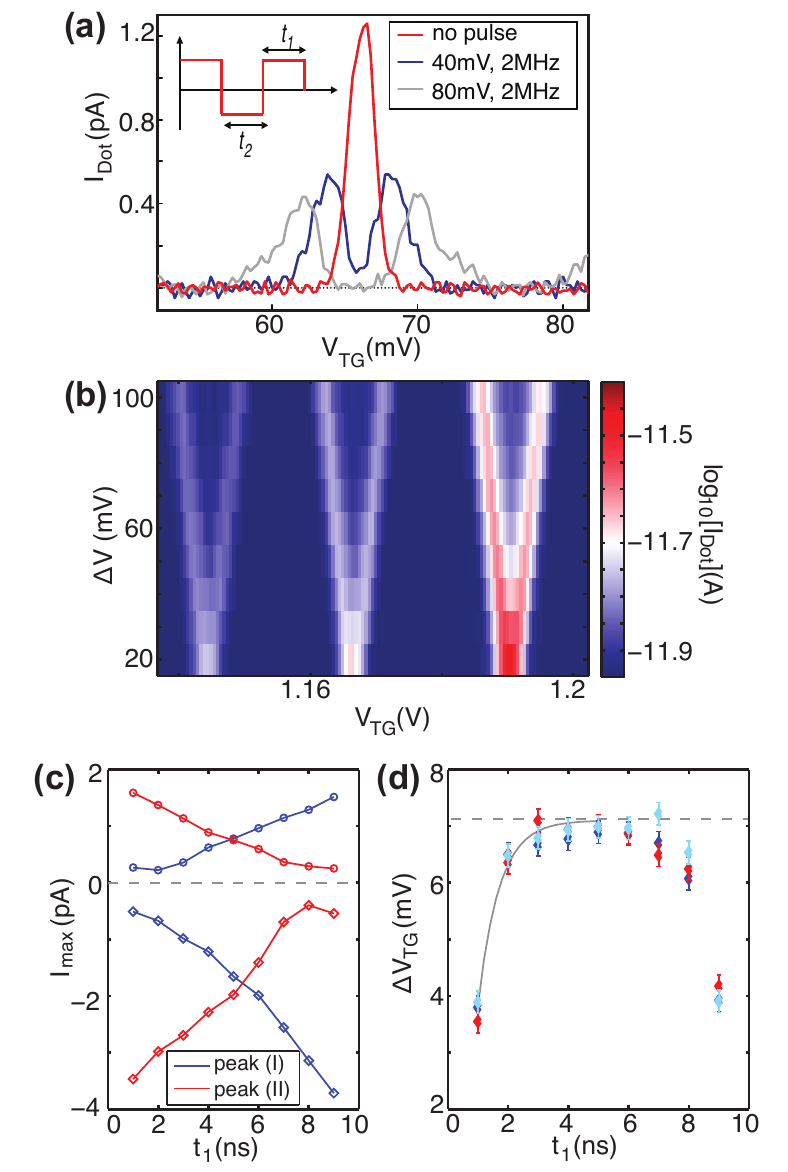}
    \caption{(a) Coulomb resonance for different amplitudes of the square pulse applied to the top gate. The inset shows the pulse shape and introduces the definition for the dwell times for two levels. (b) Evolution of three Coulomb resonances with increasing pulse amplitude $\Delta V$ for $t_\mathrm{1}$ = 5 ns ($f$ = 100 MHz). For both measurements a voltage bias of $V_\mathrm{bias}$ = 1 mV was applied. (c) Peak amplitudes for Coulomb resonance split by a $\Delta V$ = 80 mV voltage pulse with modulation frequency $f$ = 100 MHz extracted for $V_\mathrm{bias}$ = -0.5 mV (diamonds) and $V_\mathrm{bias}$ = 0.5 mV (circles). (d) Determined peak spacing as a function of pulse shape for the same Coulomb resonances at $V_\mathrm{bias}$ = -1 mV (dark blue), -0.5 mV (red) and 0.5 mV (light blue). The grey dashed line indicates the ideally achieved $\Delta_\mathrm{peaks}$ due to -20 dB attenuation at the sample. A fit to the data assuming a single exponential function is shown as the grey solid line.}
    \label{fig3}
  \end{center}
\end{figure}

The red trace in Fig.~\ref{fig3} (a) exhibits a current peak at the gate voltage at which the resonance condition for transport through the quantum dot is fulfilled.
As a symmetric square shaped AC-signal (see inset for a sketch of the pulse shape) is superimposed to the DC top gate voltage, this Coulomb resonance splits into two peaks, since the QD level comes into resonance two times as the TG is tuned towards more positive voltages - first for the upper pulse level and then for the lower. For the blue and the grey trace, the system spends an equal amount of time in the respective voltage level and hence the newly arising peaks exhibit approximately half the height of the original resonance. A slight asymmetry can be observed for the two peaks emerging at $\Delta V$ = 80 mV. We attribute this to a smeared-out pulse shape due to the damping of higher harmonics along the transmission line.

The distance between the peak maxima increases linearly with the amplitude of the applied voltage pulse. This is seen more clearly in Fig.~\ref{fig3} (b), where the splitting of three consecutive Coulomb resonances was recorded as the pulse amplitude was increased. The splitting can be used to extract the conversion factor between the signal amplitude $\Delta V$ at the AWG-output and $\Delta_\mathrm{peaks}$ present at the structure. Here, the 100 mV provided at the top of the cryostat correspond to $\Delta_\mathrm{peaks} \approx$ 8.5 mV, which is compatible with a  -20 dB attenuation at the 1 K-pot and a small additional attenuation due to the cables and the bias-tee. This is an indication that even though the AC-modulation was carried out with a repetition rate of $f$ = 100 MHz in this measurement, the square pulse signal arrives almost unperturbed at the sample.

We explain the latter finding by considering the lumped element equivalent circuit for a transmission line to describe the HF line on the chip. This model includes segments consisting of a series inductance $L_\mathrm{line}$, a series resistance $R_\mathrm{line}$, a shunt capacitance $C_\mathrm{line}$ to the substrate and a shunt conductance $G_\mathrm{line}$ as the lumped elements for the transmission line. For the frequencies considered here the system is in the limit of large signal wavelength $\lambda \approx$ 3 m compared to the transmission line length $l\approx$ 0.5 mm and hence only one segment of such a model has to be taken into account. Additionally, typical on-chip inductances of $L_\mathrm{line}\approx$ 1 nH/mm display a small contribution to the impedance compared to the series resistance $R_\mathrm{line}\gtrsim$ 1 $\Omega$/mm of the gate electrode. The admittance, on the other hand, is mainly determined by the shunt capacitance $C_\mathrm{line}\approx$ 1.2 pF of the pulsed gate to the highly doped Si back gate and $G_\mathrm{line}$ can be neglected. The resulting equivalent circuit is hence reduced to the series resistance $R_\mathrm{line}$ shunted by the capacitance $C_\mathrm{line}$. Effectively, such an arrangement resembles a low pass filter with cut-off frequency $f_\mathrm{c}$. The reactance is determined by the rise time $\tau_\mathrm{rise}$ of the pulse generator. In the measurement setup used here, this value was $\tau_\mathrm{rise} \approx$ 1 ns resulting in $|X_\mathrm{gate}| = 1/(2\pi f_\mathrm{max}\cdot C_\mathrm{gate})\approx$ 130 $\Omega$, which is much larger than $R_\mathrm{line}$ and no considerable voltage division is expected. This is in agreement with the observed pulse shape.

Consequently, we have shown that high frequency manipulation of graphene samples having a back gate, is feasible in the limits discussed above. On the sample itself, however, series resistances can occur e.g. at the interface of metal and graphene for an in-plane gate. In that case, a contact resistance of $R_\mathrm{contact}\approx$ 1 k$\Omega$ should be taken into account. Therefore, an increased bandwidth can be achieved by avoiding the use of in-plane gates for pulsing and by minimizing the area of the gate electrode. Here, experiments in which the TG electrode was used as the pulsed gate were carried out and are discussed in the following. As an upper boundary for $R_\mathrm{line}$ we therefore find a value of $\approx$ 200 $\Omega$ for the present device.

In order to probe the time evolution of electronic states on the nanosecond time scale it is conducive to change the pulse shape, meaning that the system spends nominally more time in the lower lying energy level than in the upper, or vice versa. Since the peak amplitude resembles the contribution to transport each level makes, it is therefore reduced for the shorter populated level.

Time dependent processes can be studied by varying the pulse shape \cite{Fujisawa:2001}. In  Fig.~\ref{fig3} (c) the peak amplitude  as a function of dwell time is shown. We measured a peak pair with $V_\mathrm{bias}$ = -0.5 mV (diamond shaped markers) and with $V_\mathrm{bias}$ = 0.5 mV (circular markers) and extracted the maxima for each split peak, labeled peak (I) and (II). The current amplitude depends almost linearly on $t_\mathrm{1}$ in this regime, indicating constant current flow . For the lowest dwell times, however, the current values level off and an extrapolation to 0 and 10 ns would still yield a finite current. Such a behavior is unphysical and indicates a constraint given by the setup.

To investigate this fact further we analyze the peak spacing as well and find it to be reduced for the shortest dwell times ($t_\mathrm{1}$ = $t_\mathrm{2}$ = 1 ns). An analysis of the peak spacing as a function of the parameter $t_\mathrm{1}$ is shown in Fig.~\ref{fig3} (d). The distance between three peak pairs was investigated for this graph all being excited by a 80 mV pulse amplitude. Between $t_\mathrm{1}$ = 2 ns and 8 ns the desired splitting $\Delta_\mathrm{peaks}\approx$ 7.1 mV (marked by the grey dashed line) is achieved satisfactorily. As the dwell time for one voltage level is less than 2 ns, however, the amplitude breaks down indicating the limited pulse quality. Fitting the saturation curve to an exponential function of the form ($\Delta_\mathrm{peaks,ideal}-\alpha\cdot$exp($-t_\mathrm{1}/ t_\mathrm{rise}$)), with the free parameters $\alpha$ and $t_\mathrm{rise}$, enables us to determine the rise time of the system to be $t_\mathrm{rise}\approx$ 0.7 ns. Since the rise time specified for the AWG is $<$ 1.5 ns, the value found here can well be explained by this instrument limitation and is not determined by the device or the setup itself. Due to the reduced peak separation, an overlap between split peaks is likely which may contribute to the previously observed current saturation for low dwell times.

Measurements summarizing the observations made in this Letter are shown in Fig.~\ref{fig2} (b) and (c). The finite-bias spectroscopy was carried out both without and with AC-modulation of the top gate. The absence of charge rearrangements indicates that potential fluctuations in the environment are not relevant even with the high frequency manipulation of the applied gate voltage. For the measurements in which the gate is pulsed a doubling of the Coulomb diamonds is observed. In Fig.~\ref{fig2} (c), the two newly arising peaks are visible at the touching points of the small and the big diamonds.

Resonances are observed outside the region of suppressed current running parallel to the edges of the Coulomb diamonds in Fig.~\ref{fig2} (b) and (c), which likely display the single particle spectrum of the quantum dot. The transient transport scheme introduced in Ref. \onlinecite{Fujisawa:2001} did however not reveal signatures for the relaxation of electrons from an excited state to the corresponding ground state. This may be explained by a relaxation rate $W$ much larger than the tunneling rate $\Gamma_\mathrm{drain}$ to the drain, effectively leading to transport via the ground state. Additionally, the tunneling rate from the source $\Gamma_\mathrm{source}$ should exceed $\Gamma_\mathrm{drain}$ to assure the injection of electrons into the dot at a fast rate. In to date graphene samples, the tunneling barriers exhibit resonances and can therefore not be tuned monotonously. 

In conclusion, we have shown measurement of an etched bilayer graphene quantum dot. A top gate finger, located above the island enabled us to tune the dot levels. We found the dielectric material Cytop$^{\mathrm{TM}}$ to be exceptionally stable also at cryogenic temperatures.
The high frequency measurements presented here demonstrate the possibility of pulsed gate experiments on graphene nanostructures exhibiting both a back and a top gate. One of the future prospects is to use transient current spectroscopy \cite{Fujisawa:2001} to determine coherence times of electronic states. In order to accomplish this goal, either the probing frequencies need to be increased or the characteristic time scales for the state dynamics need to be lowered. 
To overcome the latter issue, the realization of highly tunable tunneling barriers that allow for the adjustment of the decay rates to source and drain 
is desirable.

We thank P. Roulleau, C. Stampfer, F. Libisch and D. Bischoff for helpful discussions. This research was supported by the Swiss National Science Foundation through the National Centre of Competence in Research 'Quantum Science and Technology'.


\begin{thebibliography}{99}

\bibitem{Loss:1998}
D. Loss and D. P. DiVincenzo, Phys. Rev. A \textbf{57}, 120 (1998).

\bibitem{Trauzettel:2007}
B. Trauzettel, D. V. Bulaev, D. Loss, and G. Burkard, Nat. Phys. \textbf{3}, 192 (2007).

\bibitem{Schnez:2009}
S. Schnez, F. Molitor, C. Stampfer, J. G¬uttinger, I. Shorubalko, T. Ihn, and K. Ensslin, Appl. Phys. Lett. \textbf{94}, 012107 (2009).

\bibitem{Moser:2009}
J. Moser and A. Bachtold, Appl. Phys. Lett. \textbf{95}, 173506 (2009).

\bibitem{Liu:2010}
X. L. Liu, D. Hug, and L. M. K. Vandersypen, Nano Lett. \textbf{10}, 1623 (2010).

\bibitem{Guttinger:2010}
J. G\"uttinger, T. Frey, C. Stampfer, T. Ihn, and K. Ensslin, Phys. Rev. Lett. \textbf{105}, 116801 (2010).

\bibitem{Guttinger:2009}
J. G\"uttinger, C. Stampfer, T. Frey, T. Ihn, and K. Ensslin, Phys. Status Solidi B \textbf{246}, 2553 (2009).

\bibitem{Kalb:2007}
W. L. Kalb, T. Mathis, S. Hass, A. F. Stassen, and B. Batlogg, Appl. Phys. Lett. \textbf{90}, 092104 (2007).

\bibitem{Walser:2009}
M. P. Walser, W. L. Kalb, T. Mathis, T. J. Brenner, and B. Batlogg, Appl. Phys. Lett. \textbf{94}, 053303 (2009).

\bibitem{Kouwenhoven:1997}
See, e.g. L. P. Kouwenhoven, C. M. Marcus, P. L. McEuen, S. Tarucha, R. M. Westervelt, and N. S. Wingreen, in Mesoscopic Electron Transport (Plenum, New York, 1997).

\bibitem{Molitor:2009}
F. Molitor, S. Dr\"oscher, J. G\"uttinger, A. Jacobsen, C. Stampfer, T. Ihn, and K. Ensslin, Appl. Phys. Lett. \textbf{94}, 222107 (2009).

\bibitem{Libisch:2010}
F. Libisch, S. Rotter, J. G\"uttinger, C. Stampfer, and B. Burgd\"orfer, Phys. Rev. B \textbf{81}, 245411 (2010).

\bibitem{Libisch:2011}
private communication with F. Libisch.

\bibitem{Fujisawa:2001}
T. Fujisawa, Y. Tokura, and Y. Hirayama, Phys. Rev. B \textbf{63}, 081304 (2001).

\end{thebibliography}
\end{document}